\documentclass[%
 amsmath,amssymb,
 aps,
prd,showpacs,onecolumn,superscriptaddress
]{revtex4-2} 

\usepackage{graphicx} 
\usepackage{subfigure} 
\usepackage{dcolumn} 

\usepackage{amsmath} 
\usepackage{amssymb} 
\usepackage{bm} 
\usepackage{color}

\usepackage[normalem]{ulem} 
\usepackage[dvipsnames]{xcolor} 
\usepackage{hyperref} 
\usepackage{bm} 
\usepackage{caption}
\usepackage{amsthm,amsmath,amssymb} 
\usepackage{mathrsfs} 
\usepackage{url}

\begin{document}

\preprint{APS/123-QED}

\title{Prospect of detecting magnetic fields from strong-magnetized binary neutron stars}

\author{RunDong Tang}
\affiliation{Shanghai Astronomical Observatory, Shanghai, 200030, China} 
\affiliation{School of Astronomy and Space Science, University of Chinese Academy of Sciences,Beijing, 100049, China}

\author{Wen-Biao Han} %
\email{wbhan@shao.ac.cn} 
\affiliation{Shanghai Astronomical Observatory, Shanghai, 200030, China}
\affiliation{School of Astronomy and Space Science, University of Chinese Academy of Sciences,Beijing, 100049, China}
\affiliation{School of Fundamental Physics and Mathematical Sciences, Hangzhou Institute for Advanced Study, UCAS, Hangzhou 310024, China}

\author{XingYu Zhong}
\affiliation{Shanghai Astronomical Observatory, Shanghai, 200030, China} 
\affiliation{School of Astronomy and Space Science, University of Chinese Academy of Sciences,Beijing, 100049, China}

\author{Ye Jiang}
\affiliation{Shanghai Astronomical Observatory, Shanghai, 200030, China} 
\affiliation{School of Astronomy and Space Science, University of Chinese Academy of Sciences,Beijing, 100049, China}

\author{Ping Shen}
\affiliation{Shanghai Astronomical Observatory, Shanghai, 200030, China} 
\affiliation{School of Astronomy and Space Science, University of Chinese Academy of Sciences,Beijing, 100049, China}

\author{Yu Wang}
\affiliation{Shanghai Astronomical Observatory, Shanghai, 200030, China} 
\affiliation{Guangxi Key Laboratory for Relativistic Astrophysics, School of Physical Science and Technology,
Guangxi University, Nanning 530004, China}

\date{\today} 

\begin{abstract}
Binary neutron star mergers are unique sources of gravitational waves in multi-messenger astronomy. The inspiral phase of binary neutron stars can emit gravitational waves as chirp signals. The present waveform models of gravitational wave only considered the gravitational interaction. In this paper, we derive the waveform of the gravitational wave signal taking into account the presence of magnetic fields. We found that the electromagnetic interaction and radiation can introduce different frequency-dependent power laws for both amplitude and frequency of the gravitational wave. We show from the results of Fisher information matrix that the third-generation observation may detect magnetic dipole moments if the magnetic field is $\sim 10^{17}$\,G.
\end{abstract}
\maketitle

\section{INTRODUCTION}
In the 20th century the observation of first binary pulsar PSR B1913+16 by Russell A. Hulse and Joseph H. Taylor \cite{RevModPhys.66.699,RevModPhys.66.711} indicated an energy loss due to gravitational radiation. Later the GW150914 event \cite{abbott2016observation,abbott2016gw150914} marked the first direct detection of a gravitational-wave (GW) signal from the coalescing of two black holes and opened the era of GW astronomy. Then GW170817 and a gamma-ray burst announced the first direct observation of GWs from the merger of two neutron stars and subsequent electromagnetic radiation \cite{abbott2017gw170817,Goldstein_2017}. Moreover, due to the accompanying electromagnetic counterparts, this event offers an independent standard siren measurement of Hubble constant \cite{hotokezaka2019hubble}. The GW170817 can also afford some constraints on physics such as the nuclear coupling of light axion fields \cite{zhang2021first} and the emitting region of the gamma-rays \cite{matsumoto2019constraints}. The merger of a binary neutron star system can be divided into distinct phases: an inspiral phase where the objects gradually approach, a merger phase marked by rapid coalescence. The GW signal emitted during the inspiral phase is a source for detectors such as the Advanced LIGO/Virgo detectors \cite{collaboration2015advanced,acernese2014advanced}, KAGRA \cite{somiya2012detector} (LVK) and future  Einstein Telescope (ET) \cite{punturo2010einstein}. The Advanced LIGO/Virgo are expected to give a merger rate of BNSs ranging from $\sim 0.4$ to $\sim 400\,\rm yr^{-1}$ \cite{abadie2010predictions} and the upper limit is 12600\,$\rm Gpc^{-3} yr^{-1}$ \cite{abbott2016upper} since it has a considerable amount distribution in our Milky Way. Lately, the study of systems that emit electromagnetic radiation when they coalesce has attracted interest due to the rich information that can be extracted from this scenario.

Neutron star gains its strong magnetic field due to the conservation of magnetic flux after the collapse \cite{10.1063/1.2900262}. Previous observations provided that the magnetic field carried by neutron stars can reach up to $10^{15}$\,G \cite{ferrario2005magnetic}. In some relativistic simulations the magnetic fields can reach the value as strong as $10^{18}$\,G \cite{bocquet1995rotating,cardall2001effects}. A neutron star that contains a magnetic field can be treated as a magnetic dipole. The motion of a magnetic dipole and its precession around the axis of rotation can give rise to electromagnetic radiation \cite{pacini1967energy}. Some researches show that a constant magnetic dipole moving arbitrarily can emit electromagnetic radiation \cite{heras1994radiation}, since the moving magnetic dipole moment is equivalent to a current density vector and then the radiated potential can be derived by solving Maxwell equations \cite{ioka2000gravitational,10.1119/1.3591336,heras1994radiation}. On the other hand, when the magnetic dipole moment of a neutron star is misaligned with is spin axis, spin-down will take place due to the energy loss \cite{dall2011gamma}. Such neutron stars with changing magnetic dipole moments, observed as pulsars, can support rich observational effects in multi-messenger astronomy.

In the past, most researches focused on GW emissions by simulating large set of BNS systems before, during and after mergers. Many of these simulations aimed to measure the equations of state, explore the effects of magnetohydrodynamics on GWs during the process, or study the remnants following the merger \cite{anderson2008magnetized,andersson2011gravitational,bauswein2012measuring,BAIOTTI2019103714}. Data analyses have been conducted to test general relativity, particularly after the detection of the GW170817 events \cite{abbott2019tests,radice2019multimessenger,abbott2019properties,abbott2020gw190425,abbott2021gwtc}. The detection of GW from inspiral phase demands accurate waveform templates owing to the weak amplitude of wave. For the detection of BNSs, Most of the present gravitational waveform models considered only the gravitational interaction containing the effects of tidal deformation and spin \cite{dietrich2017closed,apostolatos1994spin}. Few researches focused on the electromagnetic interaction in detail, which we will introduce and examine. Due to the presence of strong magnetic fields, both the electromagnetic dipolar interaction and energy loss can affect the evolution of orbital separation distance so that the time-evolution of orbital quadrupole moments differs from that in a purely gravitational dominant system \cite{vasuth2003gravitational}. Energy loss of a magnetized neutron star originates not only from the GWs emission but also from the electromagnetic waves (EMWs). It is certain that the presence of magnetic dipole moments can affect the amplitude and changing rate of angular frequency of GWs, see Ref. \cite{ioka2000gravitational} \cite{bourgoin2022impact}. Additionally, the relativistic simulations in Refs. \cite{bocquet1995rotating} and \cite{cardall2001effects} provides a possibility of the presence of strong magnetic fields, starting from which we deduce the frequency-domain waveform model containing the effect of magnetic dipole moments and evaluate the possibility of detection.

This article is organized as follows. In Sec. \ref{section:eom} we derive the equation of motion for two magnetized neutron stars from the circular-orbit case using the Euler-Lagrange equation. We adopt the post-Newtonian method including dipolar interaction. In Sec. \ref{section:inspiral} we calculate the total energy loss rate contributed by both gravitational radiation and electromagnetic radiation averaged over a period in the adiabatic approximation. We use this result to compute the time derivative of orbital radius and further derive the waveform governed by gravitational and electromagnetic interaction by performing Fourier transform and stationary phase approximation in Sec. \ref{section:waveform}. We compare the waveform containing the influence of magnetic field with that predicted for a system dominated by pure gravity, using matched-filtering techniques for LIGO and ET in Sec. \ref{section:numerical}. In addition, we report in this section the results of evaluating the parameter estimation obtained from the Fisher information matrix.

\section{EQUATION OF MOTION FOR A BOUND BINARY SYSTEM}\label{section:eom}
Throughout this paper, we choose Gaussian units such that $\frac{\mu_0}{4\pi}=\frac{1}{4\pi \epsilon_0}=1$ and keep gravitational constant $G$ and the speed of light $c$ within the expressions. For a binary system formed with a long initial separation, it spends most of its lifetime in the inspiral phase \cite{liu2020gravitational,maggiore2007gravitational,chu2022formation} during which the lowest order post-Newtonian approximation is employed to describe it. Thus we consider the Keplerian orbits of two magnetized neutron stars with masses $m_1$ and $m_2$ and magnetic dipoles moments $\vec{d_1}$ and $\vec{d}_{2}$, respectively. For simplicity, we assume that the magnetic dipole moments of these two neutron stars are aligned with their respective spin axes and with the angular momenta of orbits.

It is necessary to introduce the relationship between magnetic dipole moment and the surface magnetic field $B$ at pole: $d\equiv |\vec{d}|=\frac{1}{2}R_{N}^{3}B$, where $R_{N}$ is the radius of a neutron star and we set the same radius for the two neutron stars throughout this paper. Moreover, we neglect the influence of spin on the orbital motions. Due to electromagnetic interaction, the orbits will differ from those of a binary system dominated solely by gravity.

Given a dipole $\vec d_2$ immersed in the field generated by a dipole $\vec d_1$, the interaction potential energy is, according to field theory \cite{Landau}:
\begin{align}
    U_D=\frac{\vec{d_1} \cdot \vec{d}_{2}-3(\vec{d_1} \cdot \vec{R}_{\mathrm{u}})(\vec{d_2} \cdot \vec{R}_{\mathrm{u}})}{R^3},
\end{align}
here we have defined $R=|\vec{R}|$, $\vec{R}=\vec{r}_{1}-\vec{r}_{2}$  and $\vec{R}_{\mathrm{u}}=\vec{R}/{R}$. In our case this expression is simplified as
\begin{align}
    U_D=\frac{d_{1}d_{2}}{R^3},
\end{align}

The Lagrangian for the bound system is given by
\begin{align}
    L=\frac{1}{2}m_{1}\dot{r}_{1}^{2}+\frac{1}{2}m_{1}\dot{r}_{2}^{2}+\frac{Gm_{1}m_{2}}{R}-\frac{d_{1}d_{2}}{R^3}.
\end{align}
The dot above denotes the derivation with respect to time $t$. It is convenient to choose the center of mass as the origin of coordinates, i.e.,
\begin{align}
    \vec{r}_{CM}=\sum_{i=1}^{2}\frac{m_{i}\vec{r}_i}{M}=(0,0),
\end{align}
where, $M=m_{1}+m_{2}$ is the total mass of the system. Therefore, $\vec{r}_{1}=m_{2}\vec{R}/M$ and $\vec{r}_{2}=-m_{1}\vec{R}/M$. By introducing the reduced mass $\mu=m_{1}m_{2}/M$ and transforming to polar coordinates $(r,\varphi)$, we can rewrite the Lagrangian as
\begin{align}\label{eq:lagrangian}
    L=\frac{1}{2}\mu\dot{R}^{2}+\frac{1}{2}\mu R^{2}\dot{\varphi}^2+\frac{G\mu M}{R}-\frac{d_{1}d_{2}}{R^3}=T-U.
\end{align}
According to Eq. (\ref{eq:lagrangian}), we find that the respective kinetic and potential energies are:
\begin{align}
    T=\frac{1}{2}\mu\dot{R}^{2}+\frac{1}{2}\mu R^{2}\dot{\varphi}^2,
\end{align}
\begin{align}
    U=-\frac{G\mu M}{R}+\frac{d_{1}d_{2}}{R^3}.
\end{align}
It is clear that the system has 2 degrees of freedom and the orbital motion of the system takes place in a two-dimensional space. We can express the Euler-Lagrange equation in terms of the $\varphi$ dimension as:
\begin{align}
    \frac{d}{dt}\left(\frac{\partial L}{\partial \dot{\varphi}}\right)-\frac{\partial L}{\partial \varphi}=\frac{d}{dt}\left(\mu R^{2}\dot{\varphi}\right)=0.
\end{align}
Furthermore, the canonical momentum $P_{\varphi}$ is given by
\begin{align}
    P_{\varphi}=\frac{\partial L}{\partial \dot{\varphi}}=\mu R^{2}\dot{\varphi}.
\end{align}
As a result, the canonical momentum $P_{\varphi}$, commonly referred to as the orbital angular momentum $l$, is conserved during the whole process. This gives the relationship between angular velocity and angular momentum: 
\begin{align}
    \dot{\varphi}=\frac{l}{\mu R^2}.
\end{align}
Subsequently, if we examine the Euler-Lagrange equation for $R$, we obtain:
\begin{align}\label{eq:radial motion}
    \frac{d}{dt}\left(\frac{\partial L}{\partial \dot{R}}\right)-\frac{\partial L}{\partial R}=\mu \ddot{R}-(\frac{l^2}{\mu R^3}-\frac{G\mu M}{R^2}+\frac{3d_{1}d_{2}}{R^4})=0.
\end{align}
Multiplying by a factor $\dot{R}$ on both sides of Eq. (\ref{eq:radial motion}) we find
\begin{align}
    \frac{d}{dt}\left(\frac{1}{2}\mu \dot{R}^{2}+\frac{l^2}{2\mu R^2}-\frac{G\mu M}{R}+\frac{d_{1}d_{2}}{R^3}\right)=0,
\end{align}
and so the total energy of this system
\begin{align}
    E=\frac{1}{2}\mu \dot{R}^{2}+\frac{l^2}{2\mu R^2}-\frac{G\mu M}{R}+\frac{d_{1}d_{2}}{R^3}
\end{align}
is conserved, as expected. Under the circular condition, both the radial velocity and acceleration vanish, leading to a simplified equation of motion:
\begin{align}\label{eq:eom}
    \mu \omega_{s}^{2}R=\frac{G\mu M}{R}-\frac{3d_{1}d_{2}}{R^4},
\end{align}
where we have defined $\dot{\varphi}\equiv\omega_{s}$, obviously it is a constant. 

\section{EVOLUTION OF ORBITAL RADIUS}\label{section:inspiral}
According to the Virial theorem, for an N-body system bounded by a potential, if the potential energy $U$ is the sum of power functions of $r$: $U=\sum a_{n}r^{n}$, the relationship between the average total kinematic energy and potential energy is
\begin{align}
    \langle T \rangle=\frac{1}{2}\langle \nabla U \cdot \vec{r}\rangle=\frac{1}{2}\sum_{n} a_{n}nr^n,
\end{align}
where $n<-1$ for BNS system such that the potential reaches zero asymptotically at infinity. Thus the kinematic energy in our case can be written as
\begin{align}
    T=\frac{1}{2}\left(\frac{G\mu M}{R}-\frac{3d_{1}d_{2}}{R^3}\right),
\end{align}
so the total energy has the form:
\begin{align}\label{eq:total energy}
    E=-\frac{G\mu M}{2R}-\frac{d_{1}d_{2}}{2R^3}=-\frac{G\mu M}{2R}\left(1+\frac{d_{1}d_{2}}{G\mu M}\frac{1}{R^2}\right).
\end{align}

\subsection{Electromagnetic radiation}
Due to the presence of both gravitational and electromagnetic radiation, the conservation of total energy is no longer maintained. We will start by investigating the energy emission resulting from electromagnetic radiation. As the neutron star is moving, it is necessary to consider the transformation of electromagnetic fields in two reference frames. Denoting the rest frame of neutron star by $S'$ and observer frame by $S$, the magnetic dipole moment $\vec{d}'$ in $S'$ is static. When $S'$ is moving with arbitrary velocity $\vec v$, it will induce an electric dipole moment $\vec{d}_{e}$ according to Lorentz Transformation of 4-potential 
$A_\mu$. In the low-velocity limit, the electric dipole moment can be expressed as, in the frame $S$ \cite{griffiths2005introduction},
\begin{align}
    \vec{d}_{e}=\frac{\vec v\times \vec{d}'}{c^2}.
\end{align}

On the other hand, the radiated field $\Vec{B}_{R}$ at field point $\vec{r}$ of a magnetic dipole in arbitrary motion $\vec{s}(t')$ is calculated by Heras \cite{heras1994radiation}: 
\begin{align}
    \Vec{B}_{R}&=\frac{3\vec{D}_{\mathrm{u}}\times (\vec{D}_{\mathrm{u}}\times \vec{d}'-c\vec{d}_{e})(\vec{D}_{\mathrm{u}}\cdot \vec a)^2}{DK^{5}c^4}+\frac{3\vec{D}_{\mathrm{u}}\times (\vec{D}_{\mathrm{u}}\times \dot{\vec{d}}'-c\dot{\vec{d}}_{e})(\vec{D}_{\mathrm{u}}\cdot \vec a)}{DK^{4}c^3}\nonumber\\
    &+\frac{\vec{D}_{\mathrm{u}}\times (\vec{D}_{\mathrm{u}}\times \vec{d}'-c\vec{d}_{e})(\vec{D}_{\mathrm{u}}\cdot \dot{\vec{a}})}{DK^{4}c^3}+\frac{\vec{D}_{\mathrm{u}}\times (\vec{D}_{\mathrm{u}}\times \ddot{\vec{d}}'-c\ddot{\vec{d}}_{e})}{DK^{3}c^2},
\end{align}
where $D=|\vec{r}-\vec{s}(t')|$, $\vec{D}_{\mathrm{u}}=(\vec{r}-\vec{s}(t'))/D$, $K=1-\vec{v}\cdot \vec{D}_{\mathrm{u}}/c$ and $\vec{a}=\dot{\vec{v}}$.
While this expression is complicated. Under the low-velocity limit and if we consider the source is small in comparison with the distance i.e., $|\vec{s}(t')|/D<<1$, the calculation can be performed up to the leading order of $\vec v/c$ and $|\vec{s}(t')|/D$, following Ref. \cite{ioka2000gravitational}: 
\begin{align}
    \vec{B}_{R}=\frac{1}{D}(\vec{D}_{\mathrm{u}}\cdot \vec{d}')[(\vec{D}_{\mathrm{u}}\cdot \dot{\vec{a}})\vec{D}_{\mathrm{u}}-\dot{\vec{a}}].
\end{align}
Then we can calculate the total power of electromagnetic wave by integrating over the solid angle \cite{Landau}: 
\begin{align}
    P_{EM}=\int \frac{dP_{EM}}{d\Omega}d\Omega=\int \frac{cB^{2}_{R}D^2}{4\pi}d\Omega.
\end{align}
The only well-defined energy emission power is given by the time average over one period of the wave:
\begin{align}
    \overline{P}_{EM}=\frac{1}{T}\int_{0}^{T} dtP_{EM}.
\end{align}
So the energy emission power averaged over a period is given by
\begin{align}
\langle \frac{dE_{EM}}{dt}\rangle=-\overline{P}_{EM}=\frac{4d^{2}R^{2}\omega_{s}^{6}}{15c^{2}},
\end{align}
where $d$ is the modulus of the effective magnetic dipole moment $\vec{d}=(m_{2}\vec{d}_1-m_{1}\vec{d}_2)/M$.

\subsection{Gravitational radiation}
To continue, we need to calculate the energy emission by the gravitational radiation. Following Ref. \cite{maggiore2007gravitational}, the total radiated power in the quadrupole approximation integrated over all the directions is (we have employed the Einstein summation convention)
\begin{align}
    P_{GW}=\frac{G}{5c^5}\langle \dddot{M}_{ij}\dddot{M}_{ij}-\frac{1}{3}(\dddot{M}_{kk})^2 \rangle,
\end{align}
with the mass quadrupole moment in the equatorial plane:
\begin{align}
    M_{ij}=\mu x^{i}x^{j}.
\end{align}

We are working in the case of a circular orbit and slow enough rate of emission of GW, so the radius of the binary and the time derivative of $\varphi$ remain constant instantaneously. According to the relation between Cartesian coordinates and polar coordinates, we finally derive the specific expressions of non-zero components of mass quadrupole moments:
\begin{align}
    &\dddot{M}_{11}=\frac{4(GM)^{3/2}\mu^{5/2}}{R^{7/2}}(1-\frac{3d_{1}d_{2}}{G\mu M}\frac{1}{R^2})^{3/2}\sin2\varphi,\nonumber\\
    &\dddot{M}_{12}=\dddot{M}_{21}=\frac{-4(GM)^{3/2}\mu^{5/2}}{R^{7/2}}(1-\frac{3d_{1}d_{2}}{G\mu M}\frac{1}{R^2})^{3/2}\cos2\varphi,\nonumber\\
    &\dddot{M}_{22}=\frac{-4(GM)^{3/2}\mu^{5/2}}{R^{7/2}}(1-\frac{3d_{1}d_{2}}{G\mu M}\frac{1}{R^2})^{3/2}\sin2\varphi.
\end{align}
Note that the emission power of GWs averaged over one period is the same as the instantaneous power:
\begin{align}
    \overline{P}_{GW}=P_{GW}.
\end{align}
Therefore, the energy emission power of GWs averaged over one period is
\begin{align}
    \langle\frac{dE_{GW}}{dt}\rangle=-\overline{P}_{GW}=\frac{32G\mu^{2}R^{4}\omega_{s}^{6}}{5c^5}.
\end{align}
and finally the rate of total energy emission is 
\begin{align}
    \langle \frac{dE}{dt}\rangle=\langle \frac{dE_{EM}}{dt} \rangle+\langle \frac{dE_{GW}}{dt} \rangle.
\end{align}

\subsection{Evolution of orbital separation}
Due to the loss of energy, the separation distance of the binary is reducing and the two neutron stars merge. By performing time derivatives on both sides of Eq. (\ref{eq:total energy}) gives the relation between time evolution of $R$ and energy loss rate. Throughout this paper we regard the electromagnetic interaction as a perturbation of orbits, i.e., $\frac{d_{1}d_{2}}{Gm_{1}m_{2}}\frac{1}{R^2}<<1$. Further more, from Eq. (\ref{eq:eom}) we have
\begin{align}\label{EQ:omega2}
    \omega_{s}^{2}=\frac{GM}{R^3}-\frac{3d_{1}d_{2}}{\mu R^4}=\frac{GM}{R^3}(1-\frac{3d_{1}d_{2}}{G\mu M}\frac{1}{R^2}),
\end{align}
then we derive $\dot{R}$ to the first order of $d_{1}d_{2}/(G\mu MR^2)$:
\begin{align}\label{eq:dR/dt}
    \frac{dR}{dt}=-\frac{64G^{3}\mu M^{2}}{5c^{5}R^{3}}\left(1-\frac{12d_{1}d_{2}}{G\mu M}\frac{1}{R^2}+\frac{d^2}{24G\mu^2}\frac{1}{R^2}\right).
\end{align}

To continue our analysis, it is convenient to introduce dimensionless expressions. It is needed to introduce the characteristic radius:
\begin{align}
    R_{*}^{3}=\left(\frac{2GM}{c^2}\right)^{2}\left(\frac{G\mu}{c^2}\right).
\end{align}
We can therefore rewrite quantities such as $t, R, M$ as a dimensionless expression $\hat{t}=ct/R_{*}, \hat{R}=R/R_{*}, \hat{\mathcal{M}}=G\mathcal{M}/(c^{2}R_{*})$, where $\mathcal{M}$ represents any mass term such as the total mass and the reduced mass and the hat symbol above denotes the dimensionless expression. Note that, $d_{i}/(\sqrt{G}m_{i})(i=1,2)$ has the dimension of $[R]^{-1}$, so the dimensionless expression is $\hat d_{i}=d_{i}/(\sqrt{G}m_{i}R_*)$. Then Eq. (\ref{eq:dR/dt}) can be rewritten as 
\begin{align}
    \frac{d\hat{R}}{d\hat{t}}=-\frac{16}{5}\frac{1}{\hat{R}^3}\left[1-\left(12\hat{d_1}\hat{d_2}-\frac{\hat{d}^2}{24}\right)\frac{1}{\hat{R}^2}\right]=-\frac{\beta}{\hat R^3}-\frac{\alpha}{\hat R^5},
\end{align}
where we have defined $\beta=\frac{16}{5},\alpha=\frac{16}{5}\left(\frac{\hat{d}^2}{24}-12\hat{d_1}\hat{d_2}\right)$. We can integrate the differential equation as
\begin{align}
    -\int_{\hat R_0}^{\hat R}\frac{\hat R^5}{\gamma\hat R^{2}+1}d\hat R=\alpha \hat t,
\end{align}
where $\gamma=\beta/\alpha$ and we set the initial time to be 0. This gives the solution:
\begin{align}\label{eq:R-t relation}
    f(\hat R_0)-f(\hat R)=\alpha \hat t,
\end{align}
\begin{align}\label{eq:f(R)}
    f(\hat R)=\frac{\rm ln(1+\gamma\hat R^2)}{2\gamma^3}-\frac{\hat R^2}{2\gamma^2}+\frac{\hat R^4}{4\gamma}.
\end{align}
According to estimation of the magnitude, the logarithm is much less than the power parts in the expression, so we neglect the contribution of the first term in the right hand of Eq. (\ref{eq:f(R)}), thus we approximately have
\begin{align}
    f(\hat R)\simeq -\frac{\hat R^2}{2\gamma^2}+\frac{\hat R^4}{4\gamma}.
\end{align}
In order to solve Eq. (\ref{eq:R-t relation}), we need to define the time to coalescence $\hat{\tau}$ through $\hat \tau+\hat t=\hat t_0=\hat \tau_0$ and $\hat t_0$ is the time when the two neutron stars coalesce. So Eq. (\ref{eq:R-t relation}) becomes \cite{christiansen2021charged}:
\begin{align}
    f(\hat R_0)-f(\hat R)=\alpha (\hat \tau_0-\hat \tau).
\end{align}
From this equation we can get the relationship between $\hat R$ and $\hat \tau$:
\begin{align}
    \hat R=\sqrt{\frac{1}{\gamma}\left(1+\sqrt{1+4\alpha\gamma^{3}\hat \tau}\right)}.
\end{align}
In our analysis we considered the assumption that $\gamma\gg1$. Actually, the terms of $\hat d_i$ in $\alpha$ is much less than unit even though we choose a large value of $B$, which can be seen by substituting some actual values into the dimensionless expressions. Finally we have the approximate solution to the first order of $1/\gamma$: 
\begin{align}\label{eq:ndR}
    \hat R=(4\beta \hat \tau)^{1/4}\left[1+\frac{1}{2\gamma}(4\beta \hat \tau)^{-1/2}\right].
\end{align}
By inserting $\hat R$ into Eq. (\ref{EQ:omega2}) and expanding to the leading order we get (notice the magnitude $\hat{d_1}\hat{d_2}\sim 1/\gamma$):
\begin{align}\label{eq:ndo}
    \hat \omega_s=2^{-3/4}\hat M_{c}^{-5/8}(4\beta \hat \tau)^{-3/8}\left[1-\left(\frac{3}{2}\hat d_{1}\hat d_{2}+\frac{3}{4\gamma}\right)(4\beta \hat \tau)^{-1/2}\right].
\end{align}
here we set $\hat \omega_s=R_{*}\omega_s/c$ and define the chirp mass as $M_c=\mu^{3/5}M^{2/5}$ together with its dimensionless expression $\hat M_c\equiv GM_c/(c^{2}R_*)$.

\section{WAVEFORM OF GRAVITATIONAL WAVES}\label{section:waveform}
\begin{figure}[t]
    \centering
    \includegraphics[width=0.9\linewidth]{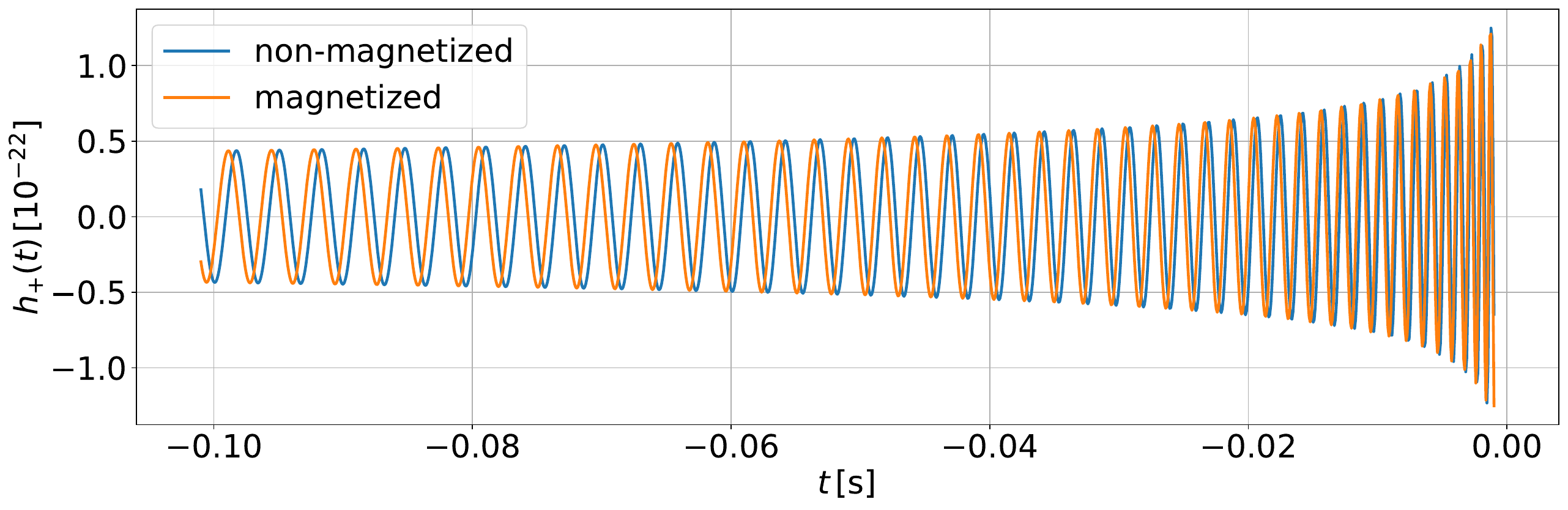}
    \caption{Time-domain waveform of $h_{+}(\hat{t})$: The reference point for time is set at the coalescence time $t_c$. We have selected $m_1=m_2=1.55\,M_{\odot}$ so that the chirp mass is $M_c\simeq 1.39\,M_\odot$. We set $R_{N}=13.8$\,km for the two neutron stars and $D=100$\,Mpc. For the magnetized neutron stars, we choose the surface magnetic field as $8\times 10^{17}\,\rm{G}$.}
    \label{fig:time domain of h+}
\end{figure}

In the theoretical analysis of waveform, we calculate the waveform in harmonic coordinates, thus there are only two independent components: the plus polarization 
\begin{align}
    h_{+}(\hat t)=\frac{4\hat \mu\hat \omega_{s}^2 \hat R^2}{\hat D}\left(\frac{1+\cos{\iota}^2}{2}\right)\cos\Phi
\end{align}
and the cross polarization
\begin{align}
    h_{\times}(\hat t)=\frac{4\hat \mu\hat \omega_{s}^2 \hat R^2}{\hat D}\cos{\iota}\sin\Phi,
\end{align}
here $\iota$ is the angle between the orbital angular momentum and the line of sight of the observer and $\hat D$ is the dimensionless luminosity distance from the binary to the observer. We select $\iota=\pi/2$ so that the GW observed has only plus polarization:
\begin{align}\label{eq:h+(t)}
    h_{+}(\hat t)=\frac{2\hat \mu\hat \omega_{s}^2 \hat R^2}{\hat D}\cos\Phi,
\end{align}
with the phase defined as:
\begin{align}
    \Phi(\hat{t})=2\int_{\hat{t}_0}^{\hat t}\hat{\omega}_{s}(\hat{t'})d\hat{t'}=\int_{\hat{t}_0}^{\hat t}\hat{\omega}_{GW}(\hat{t'})d\hat{t'}
\end{align}
where $\omega_{GW}=2\omega_{s}$ is known as the chirping frequency. From Eq. (\ref{eq:ndo}) we can see that $\omega_{GW}$ increases as the binary approaches the coalescence, i.e., $\hat{\tau}$ decreases. Substituting Eq. (\ref{eq:ndR}) and (\ref{eq:ndo}) into Eq. (\ref{eq:h+(t)}), we obtain:
\begin{align}
    h_{+}(\hat t)=A(\hat \tau)\cos\Phi,
\end{align}
with
\begin{align}
    A(\hat \tau)\simeq \frac{\sqrt{2}}{\hat D}\hat M_{c}^{5/4}(4\beta \hat \tau)^{-1/4}\left[1-\left(\frac{1}{2\gamma}+3\hat d_{1}\hat d_{2}\right)(4\beta \hat \tau)^{-1/2}\right].
\end{align}
To observe the effects of magnetic dipole moments, we plotted the waveforms of magnetized and non-magnetized neutron stars in Fig. (\ref{fig:time domain of h+}). Note that we set the magnetic field to be of order $\sim 10^{17}$, under the relativistic simulation results of Ref. \cite{cardall2001effects}.
The equations show that the magnetic dipole moments contribute additional $\hat \tau$ terms to both the frequency $\omega_{GW}(\hat \tau)$ and amplitude $A(\hat \tau)$ of wave. The additional frequency terms induce a considerable phase shift compared with the case where there are no magnetic dipole moments. Also, both the amplitude and frequency increase gradually as coalescence is approached. So the property is referred to as `chirp'.

Each detector is sensitive to signals within a specific frequency range, which varies depending on the instrument. For instance, the LIGO operates within $10\,\rm Hz-1000\,\rm Hz$ and $10^{-4}\,\rm Hz-10^{-1}\,\rm Hz$ for LISA \cite{collaboration2015advanced,k1997lisa}. In order to know the frequency distribution of a signal, it is needed to take Fourier transform and rewrite the signal waveform in the frequency domain. The expression is:
\begin{align}
    \tilde h_{+}(f)=\int_{-\infty}^{+\infty}dth_{+}(t_{r})e^{i2\pi ft_{r}}.
\end{align}
Note that the GW propagates at the speed of light $c$, when a wavefront is emitted from a source, it takes some time to reach the observer, so the integrand must be evaluated at the retarded time $t_{r}=t-D/c$. Taking into account that the differential of the retarded time is $dt_{r}=dt$ and that $\cos\Phi=\frac{1}{2}(e^{i\Phi}+e^{-i\Phi})$, we can rewrite the expression as:
\begin{align}
    \tilde h_{+}(f)=\frac{1}{2}e^{i2\pi fD/c}\int_{-\infty}^{+\infty}dt_{r}A(t_r)(e^{i\Phi(t_r)}+e^{-i\Phi(t_r)})e^{i2\pi ft_{r}}.
\end{align}
Following Ref. \cite{maggiore2007gravitational}, we use the stationary phase approximation, where only the term $e^{i(-\Phi+2\pi ft)}$ has stationary point, while the term $e^{i(\Phi+2\pi ft)}$ oscillates fast that will be integrated to a negligibly small value. Thus the expression reduces to
\begin{align}
    \tilde h_{+}(f)\simeq \frac{1}{2}e^{i2\pi fD/c}\int_{-\infty}^{+\infty}dt_{r}A(t_r)e^{i[2\pi ft_{r}-\Phi(t_r)]}.
\end{align}
At the stationary point $t_*$ we have $2\pi f=\dot \Phi(t_*)\equiv\omega_{GW}$. This indicates that the largest contribution to Fourier components is obtained for the value $t$ such that $\omega_{GW}$ is equal to $2\pi f$. Taking into account that $A(t_r)$ varies much slowly compared to phase, we expand the expression around the point $t_*$ to the second order of $(t-t_*)$ and ultimately obtain:
\begin{align}\label{eq:station h+}
    \tilde h_{+}(f)=\frac{1}{2}e^{i\Psi_+}A(t_*)\left(\frac{2}{\ddot\Phi(t_*)}\right)^{1/2},
\end{align}
where $\Psi_{+}(t_*)=2\pi f(t_*+D/c)-\Phi(t_*)-\pi/4$. From Eq. (\ref{eq:ndo}) we can immediately get, keeping it in mind that $\omega_{GW}$ is twice the angular velocity,
\begin{align}
    \Phi(\hat \tau)=\Phi(\hat \tau_0)-2^{1/4}\hat M_{c}^{-5/8}\left[\frac{8}{5}(4\beta)^{-3/8}\hat \tau^{5/8}-\left(12\hat d_{1}\hat d_{2}+\frac{6}{\gamma}\right)(4\beta)^{-7/8}\hat \tau^{1/8}\right].
\end{align}

To get the relation between $t_*$ and $f$, or equivalently between $\tau_*$ and $f$ we need to solve the equation:
\begin{align}
    \dot \Phi(\hat t_*)=\frac{d}{dt}\Phi(\hat t)|_{\hat t=\hat t_*}=-\frac{d}{d\tau}\Phi(\hat \tau)|_{\hat \tau=\hat \tau_*}=2\pi \hat f,
\end{align}
thus we derive the frequency:
\begin{align}
    \hat f=\frac{2^{-3/4}}{\pi}\hat M_{c}^{-5/8}\left(4\beta \hat \tau_*\right)^{-3/8}\left[1-\left(\frac{3}{2}\hat d_{1}\hat d_{2}+\frac{3}{4\gamma}\right)\left(4\beta \hat \tau_*\right)^{-1/2}\right].
\end{align}
Since the electromagnetic interaction is much weaker than gravity, we neglect the contribution of magnetic dipole and obtain the relation:
\begin{align}\label{eq:t of f}
    \hat \tau_*\simeq \frac{1}{16\beta}\hat M_{c}^{-5/3}(\pi\hat f)^{-8/3}.
\end{align}
Substituting Eq. (\ref{eq:t of f}) into the expressions of $A(\hat \tau_*)$ and $\ddot \Phi(\hat \tau_*)$ in Eq. (\ref{eq:station h+}) and restoring $G,c$ we finally obtained the full expression in frequency domain for further analysis:
\begin{align}\label{eq:mod waveform}
    \tilde h_{+}(f)&=e^{i\Psi_{+}}\frac{c}{D}\left(\frac{5}{96}\right)^{1/2}\nonumber\\
                   &\times\left[\pi^{-2/3}\left(\frac{GM_{c}}{c^3}\right)^{5/6}f^{-7/6}-\frac{\pi^{2/3}}{2}c^{-2}\left(\frac{GM_c}{c^3}\right)^{1/6}\left(\frac{\mu}{M}\right)^{2/5}\left(\frac{d^2}{24G\mu^2}-\frac{6d_{1}d_{2}}{Gm_{1}m_{2}}\right)f^{1/6}\right],
\end{align}
where the phase is given by 
\begin{align}
    \Psi_{+}(f)&=\Psi_{0}+2\pi f\left(\frac{D}{c}+t_0\right)\nonumber\\
               &+\frac{3}{128}\pi^{-5/3}\left(\frac{GM_c}{c^3}\right)^{-5/3}f^{-5/3}-\frac{15}{64}\pi^{-1/3}c^{-2}\left(\frac{GM_c}{c^3}\right)^{-7/3}\left(\frac{\mu}{M}\right)^{2/5}\left(\frac{d^2}{24G\mu^2}-\frac{6d_{1}d_{2}}{Gm_{1}m_{2}}\right)f^{-1/3}.
\end{align}
All the variables associated with a constant phase have been grouped together into $\Psi_0$. Moreover, since $h_+(t)$ is already dimensionless, the frequency-domain expression $\tilde h_+(f)$ carries the dimension of $[T]^{-1}$. Clearly, the alteration of magnetic dipole moments in waveform (\ref{eq:mod waveform}) introduces different dependencies on frequency in both amplitude and phase, ultimately reducing to the scenario where only gravity dominants, namely initial waveform:
\begin{align}
    \tilde h_{i,+}(f)=e^{i\Psi_{i,+}}\frac{c}{D}\left(\frac{5}{96}\right)^{1/2}\pi^{-2/3}\left(\frac{GM_{c}}{c^3}\right)^{5/6}f^{-7/6},
\end{align}
with phase
\begin{align}
    \Psi_{i,+}(f)=\Psi_{i,0}+2\pi f\left(\frac{D}{c}+t_0\right)+\frac{3}{128}\pi^{-5/3}\left(\frac{GM_c}{c^3}\right)^{-5/3}f^{-5/3}.
\end{align}
\section{DATA ANALYSIS}\label{section:numerical}
Using the waveform, we can analyse the dependence of $\tilde h_{+}(f)$ on physical properties of neutron stars, especially the surface magnetic field $B$. We also estimated how accurate the parameters will be identified in the future observations. In our analysis we choose $d_{2}=d_{1}/2000$ where the approximation $\gamma\gg1$ is satisfied.

\subsection{Matched filtering}
The matched filtering technique is used here to search for deviation between two waveforms. We introduce the inner product between two time series $a(t)$ and $b(t)$ \cite{lindblom2008model}:
\begin{align}
    (a|b)=4\mathrm{Re}\int_{0}^{+\infty} \frac{\tilde a(f) \tilde b^{\ast}(f)}{S_{n}(f)} df.
\end{align}
The tilde symbols stand for the Fourier transform and the star denotes the complex conjugation. The quantity $S_n(f)$ is the power spectral density (PSD) of noise for a particular detector.

We use this method to quantify the differences between our modified waveform and initial waveform from which we learn the influence of magnetic dipole. We calculate the maximized fitting factor or match between two signals to quantify the similarity between them: 
\begin{equation}
    \mathrm{match}=\underset{{t_s},{\phi_s}}{\mathbf{max}}\frac{(a(t)\vert b(t+t_{s})e^{i\phi_{s}})}{\sqrt{(a\vert a)(b\vert b)}},
\end{equation}
where the maximization is taken after some proper shift of $t_s$ and $\phi_s$. As the post-Newtonian approximation is applicable in the long-distance condition, the computation must be cut off well before one neutron star reaches the innermost stable circular orbits.

\begin{figure}[t]
    \begin{minipage}{0.45\linewidth}
      \centering
      \includegraphics[width=\linewidth]{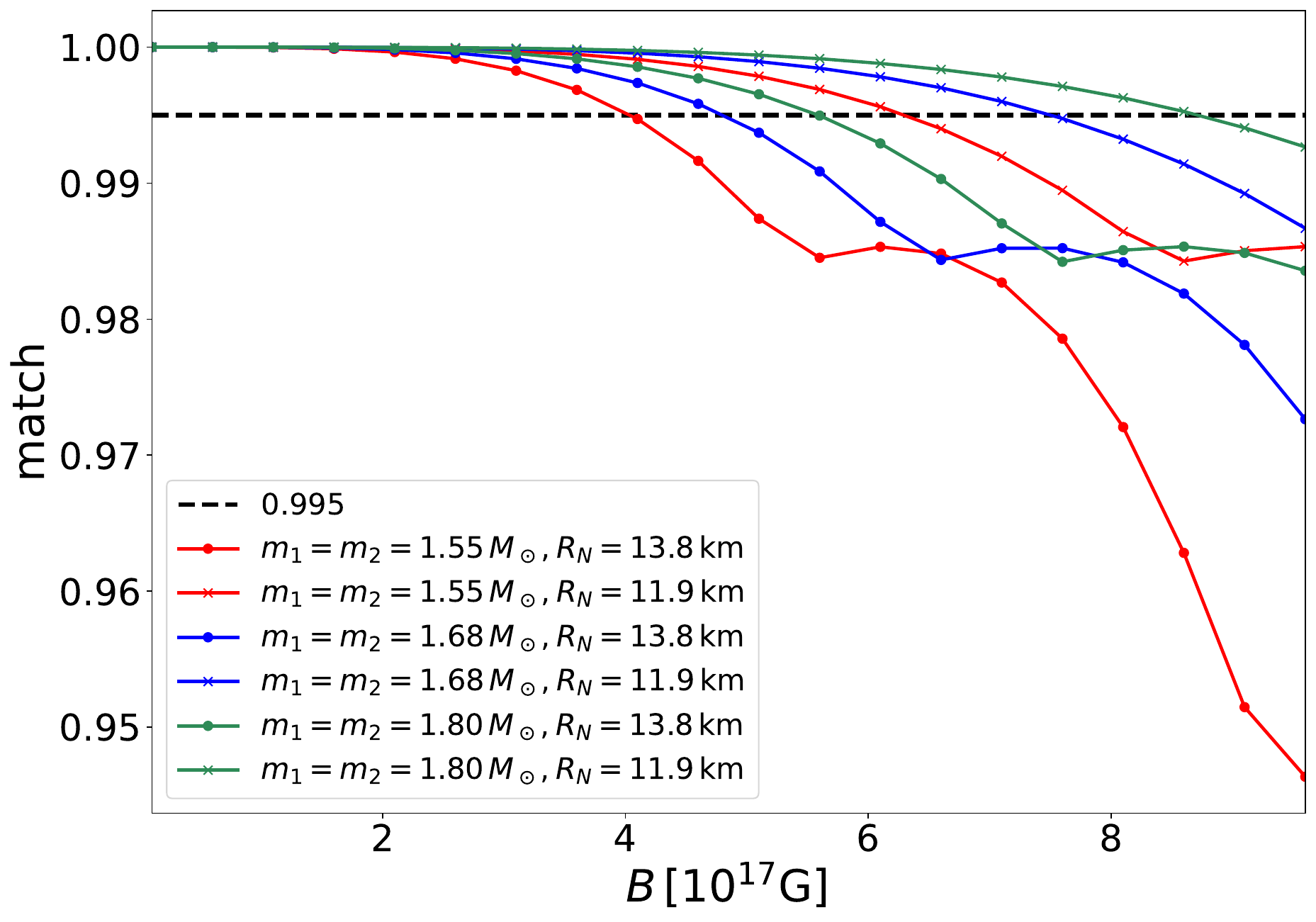}
    \end{minipage}
    \begin{minipage}{0.45\linewidth}
      \centering
      \includegraphics[width=\linewidth]{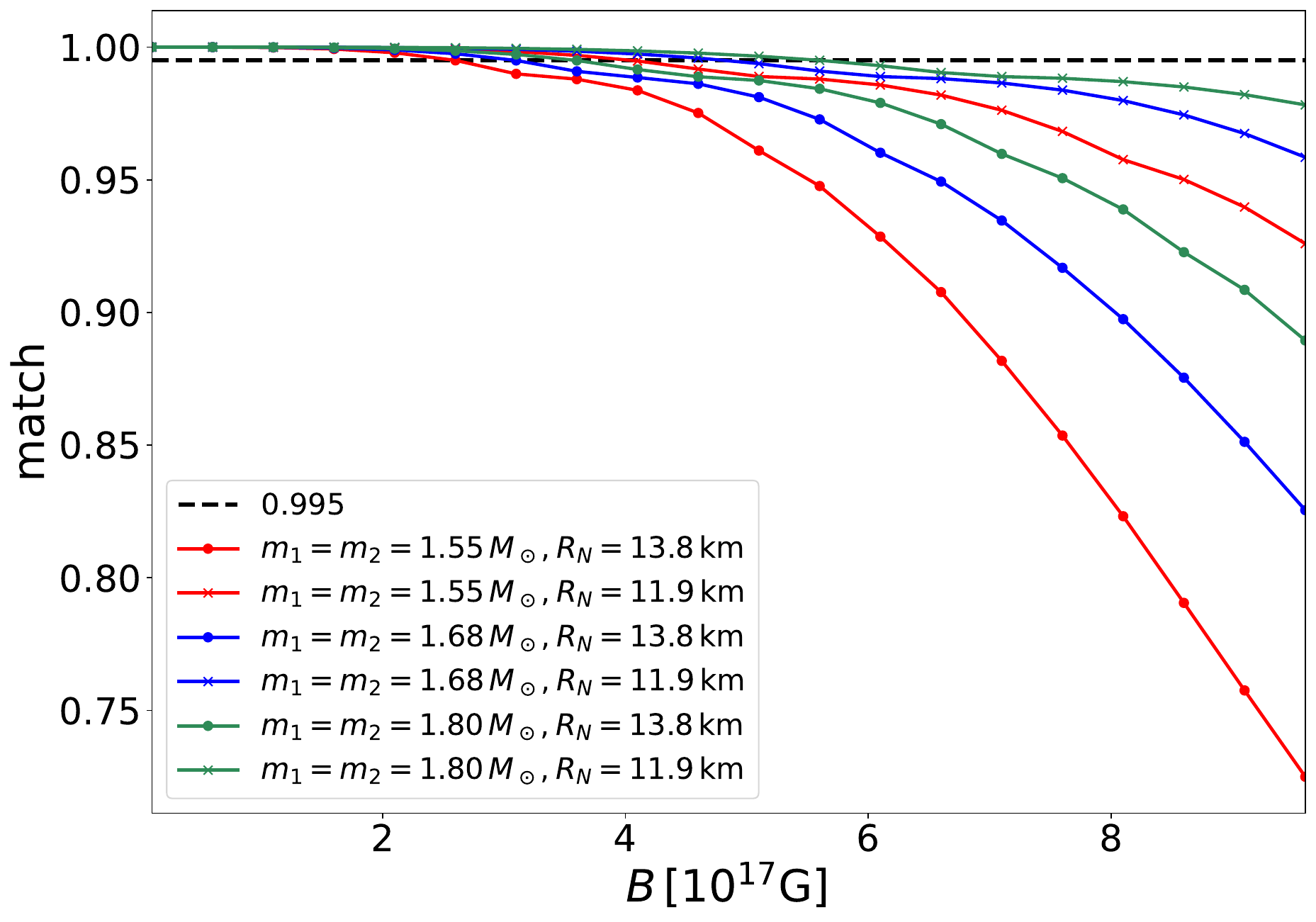}
    \end{minipage}
    \caption{Matches between the waveforms with magnetized and non-magnetized neutron stars. We set different surface magnetic fields for magnetized neutron stars. Different colors refer to different masses. The black dashed line denotes the critical value of match below which the detector is expected to well distinguish two waveforms. The left panel shows the values of match for LIGO detector and the right panel is for ET.}
    \label{fig:match}
\end{figure}

Using the relationship between magnetic dipole moment and the surface magnetic field, we compute different match for different surface magnetic fields in the frequency domain. To provide a comparison, we calculated the match for two GW detectors and illustrated the result of matched-filtering using fixed masses and radii for the neutron stars, as depicted in Fig. (\ref{fig:match}). Note that $R_N$ has same value for two neutron stars. We observe that match diminishes as masses decrease. This results from that GWs are generated due to the changing mass quadrupole moments. As we choose the electromagnetic interaction to be a perturbation of this orbital behavior, the deviation between two waveforms remains small for weak magnetic fields. 

\subsection{Evaluation of parameter estimation}
Fisher information matrix (FIM) is employed to characterize the performance in parameter estimation for detectors \cite{vallisneri2008use}. When considering the influence of parameters, the waveform is a function of parameters, i.e.
\begin{align}
    h=h(\lambda_{i}),
\end{align}
where $\lambda_i$ is the parameter vector with i ranges from 1 to the total number of parameters. Once we have the waveform Eq. (\ref{eq:mod waveform}) we can write the strain amplitude detected by a detector in the frequency domain:
\begin{align}
    h(f)=F_{+}h_{+}+F_{\times}h_{\times},
\end{align}
here the antenna pattern functions $F_+$ and $F_\times$ are given by \cite{maggiore2007gravitational}:
\begin{equation}
\begin{gathered}
    F_{+}=\frac{1}{2}(1+\cos^{2}\theta)\cos 2\phi,\\
    F_{\times}=\cos\theta \sin2\phi.
\end{gathered}
\end{equation}
where $\theta$ and $\phi$ denote the sky location. This expression has the same linear combination as that in time domain because we regard $\theta,\phi,\iota$ as time-independent. Thus, the detected strain amplitude relies on the parameters:
\begin{align}
    h&=h(f;\lambda_{i})\nonumber\\
    &=e^{i\Psi_{+}}\frac{c}{D}\left(\frac{5}{24}\right)^{1/2}\left[\frac{1}{2}(1+\cos^{2}\theta)\cos2\phi\frac{1+\cos^{2}\iota}{2}+\cos\theta \sin2\phi\cos\iota e^{i\frac{\pi}{2}}\right]\nonumber\\
   &\times \left[\pi^{-2/3}\left(\frac{GM_{c}}{c^3}\right)^{5/6}f^{-7/6}-\frac{\pi^{2/3}}{2}c^{-2}\left(\frac{GM_c}{c^3}\right)^{1/6}\left(\frac{\mu}{M}\right)^{2/5}\left(\frac{d^2}{24G\mu^2}-\frac{6d_{1}d_{2}}{Gm_{1}m_{2}}\right)f^{1/6}\right]\nonumber\\
   &=\frac{1}{4}(1+\cos^{2}\theta)\cos2\phi(1+\cos^{2}\iota)Ae^{i\Psi_+}+\cos\theta \sin2\phi\cos\iota Ae^{i\Psi_{\times}},
\end{align}
with $\Psi_{\times}=\Psi_{+}+\pi/2$. 

We are especially interested in the influence of magnetic dipole moments of neutron stars on the waveform. We would like to see how accurate we can estimate and further constrain the parameters through GWs since GWs carry information of the physical properties of neutron stars. Since the surface magnetic fields and radii of neutron stars are completely coupled in magnetic dipole moments, it is advantageous to treat the magnetic dipole moment of the first neutron star as a distinct and independent parameter.

Introducing the FIM for a given waveform $h$ \cite{vallisneri2008use}:
\begin{equation}
\begin{gathered}
    \Gamma_{ij}=\left(\frac{\partial h}{\partial \lambda_i} \Big| \frac{\partial h}{\partial \lambda_j}\right).
\end{gathered}
\end{equation}
The square root of the diagonal elements of the inverse of the FIM provides the errors of parameters: 
\begin{align}
    \Delta\lambda_i=\sqrt{(\Gamma^{-1})_{ii}}.
\end{align}
Then we can calculate the corresponding likelihood:
\begin{align}
    \mathcal L(\lambda)\propto e^{-\frac{1}{2}\sum\limits_{i,j}\Gamma_{ij}\Delta \lambda_{i}\lambda_{j}}.
\end{align}

For comparison and estimation of the accuracy, we computed relative errors of some main parameters for the two detectors. Note that we regard the magnetic dipole moment of the first neutron star $d_1$ as the only independent magnetic dipole moment. When calculating the matrix, we set the angles to be $\theta=\pi/4, \phi=\pi/3, \iota=\pi/6$ in order to contain $h_+, h_{\times}$ components. We provide the results for three main parameters in Table (\ref{tab:relative errors}).
\begin{table}[t]
    \centering
    \setlength{\tabcolsep}{20mm}
    \renewcommand{\arraystretch}{1.35}
    \large
    \begin{tabular}{ccc}\toprule
                      & LIGO & ET\\
    \hline
    $\frac{\Delta d_1}{d_1}$ & 3.52\% & 0.26\%\\
    $\frac{\Delta D}{D}$ & 4.31\% & 0.33\%\\
    $\frac{\Delta M_{c}}{M_{c}}$ & $1.04\times 10^{-5}$ & $3.13\times 10^{-7}$\\
    \botrule
    \end{tabular}
    \caption{Relative errors of $d_1$, $D$, $M_c$ are evaluated using the PSD of both LGIO and ET detectors, respectively. In the computation we employ the parameter values of $m_1=1.65\,M_{\odot}, m_2=1.55\,M_{\odot}$ and $D=100$\,Mpc. Additionally, we set $B=6.0\times 10^{17}\,\rm G$ and $R_{N}=13.8$\,km so that the magnetic dipole moment is $d_1=7.88\times 10^{35}\,\rm G\cdot \rm cm^3$.}
    \label{tab:relative errors}
\end{table}
It is important to see the lower limit that can be identified. It has been showed that a model can achieve a perfect fit to data for certain parameters when the match satisfies the condition \cite{chatziioannou2017constructing}:
\begin{align}
    1-\mathrm{match}<\frac{N}{2\rm SNR^2}
\end{align}
where $N$ is the number of intrinsic parameters (here $N=12$) of the waveforms and SNR is the signal-to-noise ratio. On the other hand, this gives the upper limit for a detector to distinguish two waveforms. From the results plotted in Fig. (\ref{fig:match}), we select $d_1=7.88\times 10^{34}\,\rm G\cdot \rm cm^3$ to compute the relative errors giving $\Delta d_1/d_1=26.21\%$. This gives an optimal estimation of the magnetic dipole moment at such a low value, particularly for a high SNR.

\begin{figure}[t]
    \begin{minipage}{0.325\linewidth}
      \centering
      \includegraphics[width=\linewidth]{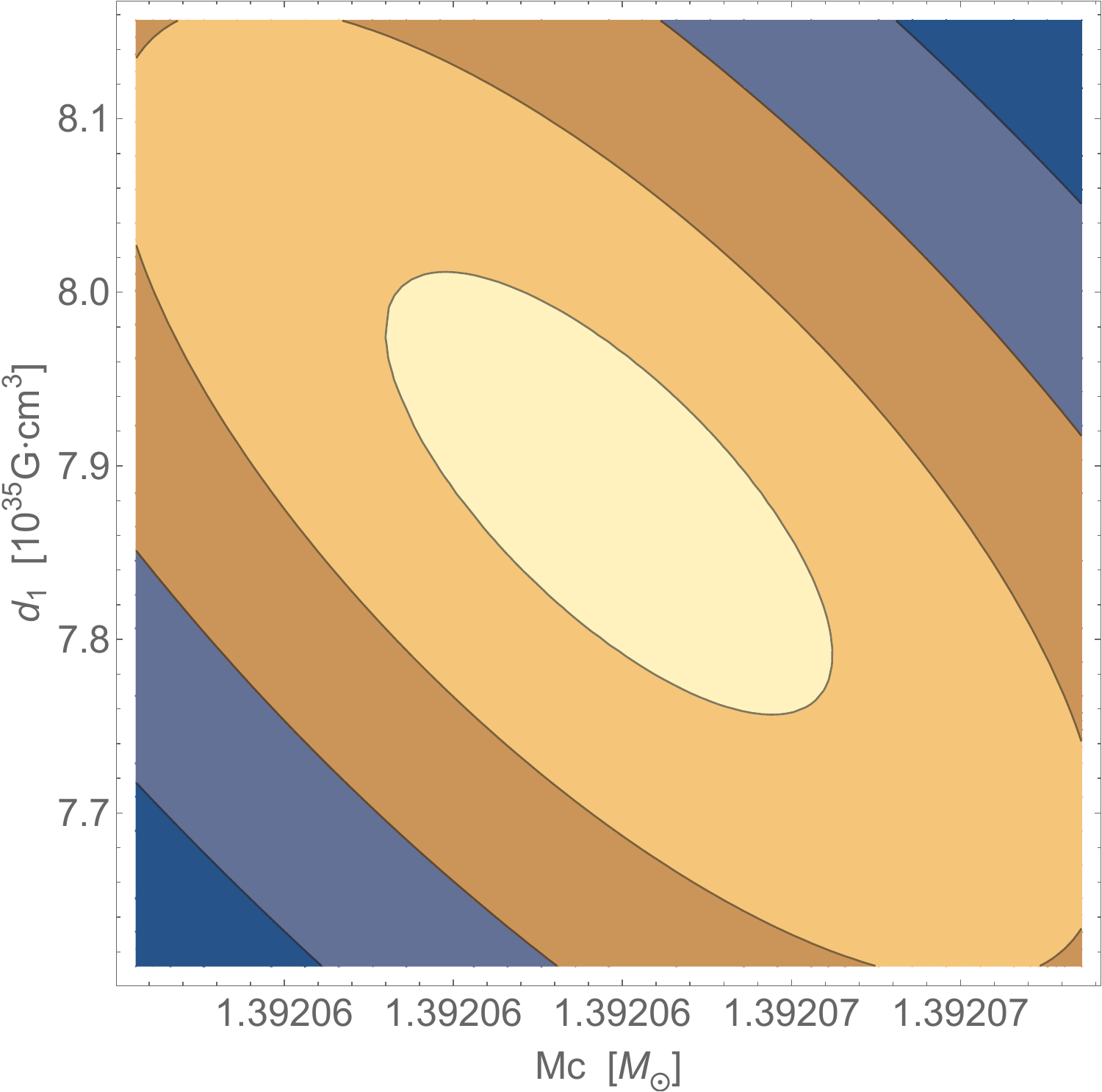}
    \end{minipage}
    \begin{minipage}{0.325\linewidth}
      \centering
      \includegraphics[width=\linewidth]{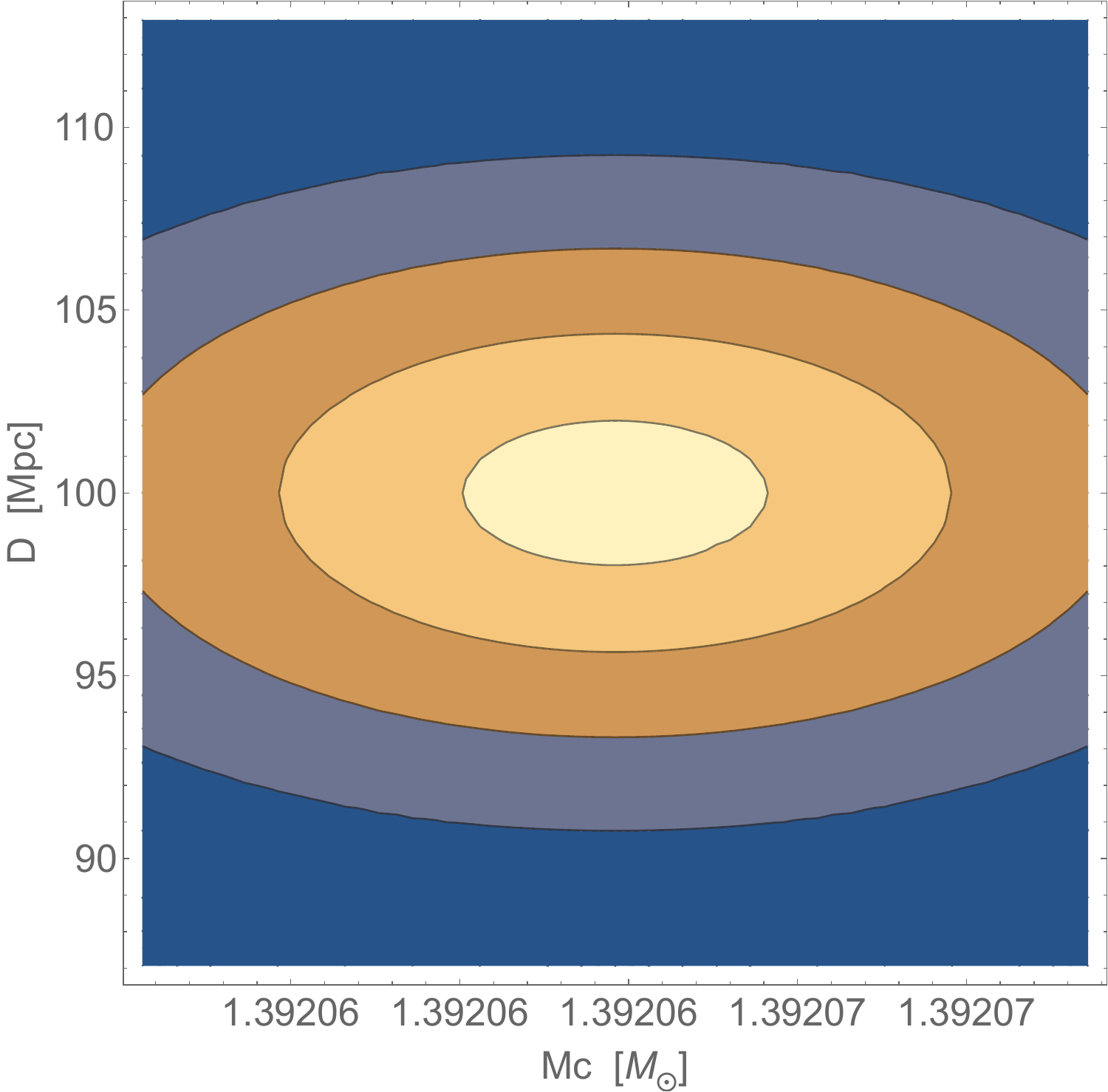}
    \end{minipage}
    \begin{minipage}{0.325\linewidth}
      \centering
      \includegraphics[width=\linewidth]{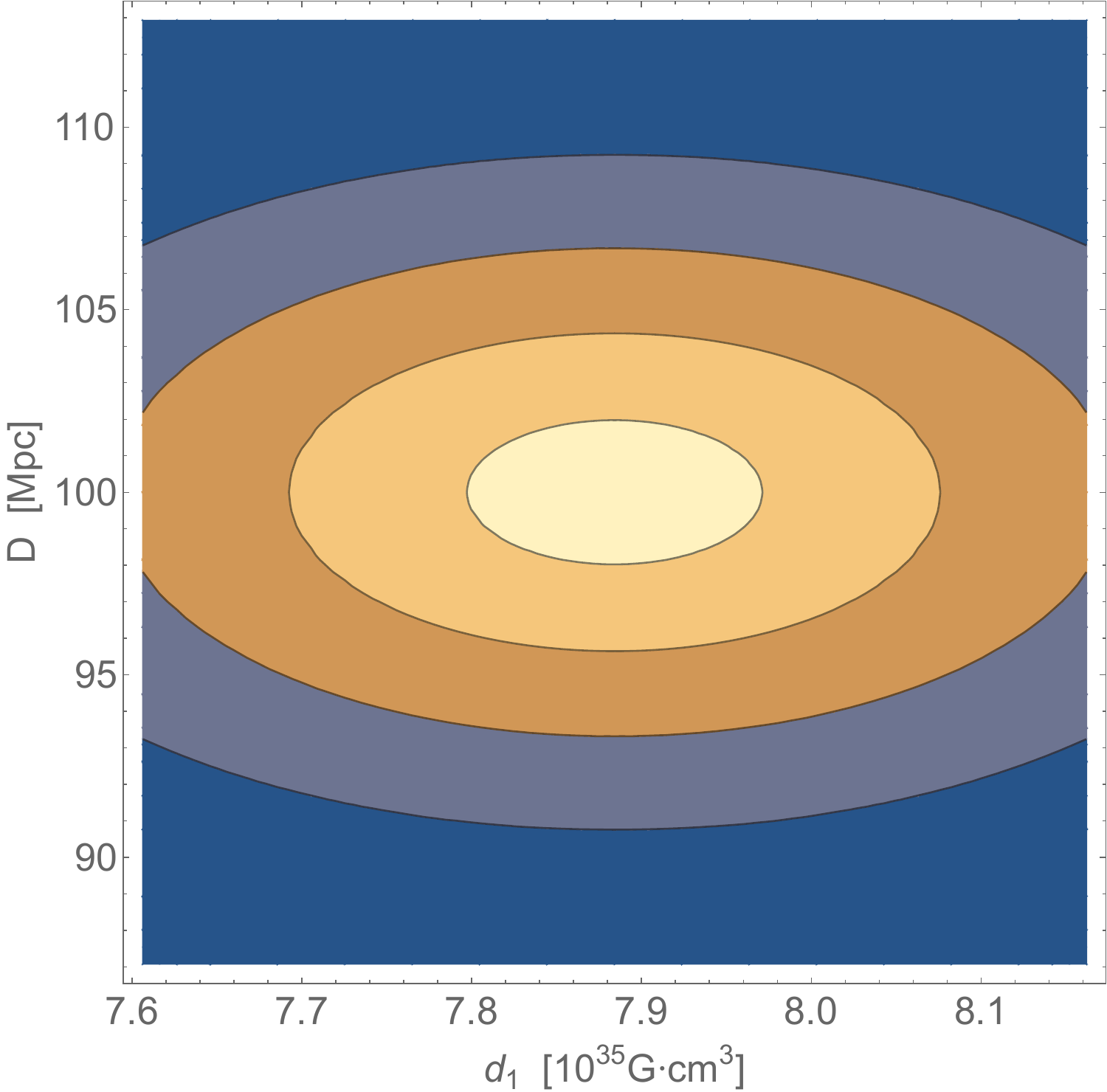}
    \end{minipage}
    \caption{Likelihood of parameters derived from FIM. The left-hand panel plotted the likelihood of chirp mass $M_c$ and the magnetic dipole moment of the first neutron star $d_1$. The figure in the middle shows the likelihood of $M_c$ and the luminosity distance from the binary to our earth $D$. The right-hand panel is for $d_1$ and $D$. In the three figures the contours from the outermost to the innermost are 10\%, 30\%, 60\% and 90\% levels respectively.}
    \label{fig:contour plot}
\end{figure}

The errors derived from the FIM show a promising evaluation in estimating the magnetic dipole moments and chirp mass, due to our selection of an extremely strong magnetic fields carried by neutron stars. We plot the contour of likelihood for LIGO, shown in Fig. (\ref{fig:contour plot}). Note that the method characterizes the performance of a detector in parameter estimation, thus this  will give a more optimistic estimation than practice.

\section{CONCLUSIONS AND DISCUSSIONS}
The mergers of binary neutron stars are promising electromagnetic counterparts of the GW sources in multi-messenger astronomy. In this paper, we have derived the equation of motion for a binary neutron star system taking into account magnetic dipolar interaction. In our analysis, we calculated the total energy emission rate and the time evolution of orbital radius for circular orbits by using the post-Newtonian method and by considering the lowest order multipole radiation for gravitational and electromagnetic waves. It is shown that both the magnetic dipolar interaction and electromagnetic radiation can modify the orbital motion and evolution. Considering the modification we calculated the gravitational waveform in the frequency domain including terms related to the magnetic dipole moment. It is found that magnetic dipole moment can introduce a significant phase shift proportional to $f^{-1/3}$ and introduce an additional term proportional to $f^{1/6}$ in the amplitude. 

After obtaining the waveforms containing magnetic dipole moment, we employ the matched filtering and the FIM for evaluating the parameter estimation for the LVK detectors and future ET detector. In the strong magnetic field regime, we show that the match of two signals with and without magnetic dipole decreases as the magnetic fields increase. This suggests the potential for detecting extremely strong magnetic fields, or provide constraints on the magnetic fields of neutron stars in future observations by the LVK and ET. The latter one could measure or constrain the magnetic field in a higher accuracy. The analysis from FIM provides an optimistic evaluation for the parameter estimations with a high signal-to-noise ratio, especially the estimation of magnetic dipole moments is limited to a few percent for ET.

Note that the improved sensitivity of future detectors will result in more detection of GWs from neutron star mergers. Since neutron stars usually carry strong magnetic fields \cite{duncan1992formation} which can affect the waveforms, this provides possibility to detect the strong magnetic fields in the future GW observation. Furthermore, we can also give a constraint on the magnitude of magnetic fields and further constrain the equations of state of neutron stars \cite{PhysRevLett.79.2176,PhysRevC.82.065802}. It is important to acknowledge that the spin of a neutron star might not align with its magnetic axis, resulting in a different modification to the dynamics in contrast to the current scenario. Thus in the future work we will generalize the current case to the magnetic dipole moments with arbitrary orientation \cite{weisberg1981gravitational}. Further the modification by magnetic dipole moments introduced in this work can be extended to the black hole-neutron star binaries. Indeed, a rotating black hole in the magnetic fields produced by the neutron star can accrete charges and form a magnetic dipole moment \cite{wald1974black,liu2016fast}, with such effects we can further investigate the black hole-neutron star mergers and give methods to search for charged black holes.

\section*{ACKNOWLEDGEMENTS}
We thank Chen Su for his discussions and suggestions for our coding work. We also thank Belahcene Imene for her valuable advice on this work. This work was supported by the National Key R\&D Program of China (Grant Nos. 2021YFC2203002), the National Natural Science Foundation of China (Grant Nos. 12173071). Wen-Biao Han was supported by the CAS Project for Young Scientists in Basic Research (Grant No. YSBR-006). This work made use of the High Performance Computing Resource in the Core Facility for Advanced Research Computing at Shanghai Astronomical Observatory.
\bibliography{reference} 

\end{document}